# Precision and Stability Issues in VBL, the Virtual Biophysics Lab simulation program

E. Milotti*, A. Del Fabbro*, and R. Chignola**

* Dipartimento di Fisica dell'Università di Trieste and INFN-Sezione di Trieste, Trieste, Italy
**** Università di Verona and INFN-Sezione di Trieste, Verona, Italy

*Abstract*—The network of biochemical reactions inside living organisms is characterized by an overwhelming complexity which stems from the sheer number of reactions and from the complicated topology of biochemical cycles. However the high speed of computers and the sophisticated computational methods that are available today are powerful tools that allow the numerical exploration of these exceedingly interesting dynamical systems. We are now developing a program, the Virtual Biophysics Lab (VBL), that simulates tumor spheroids, and which includes a reduced – but still quite complex – description of the biochemistry of individual cells, plus many diffusion processes that bring oxygen and nutrients into cells and metabolites into the environment. Each simulation step requires the integration of nonlinear differential equations that describe the individual cell's clockwork and the integration of the diffusion equations. These integrations are carried out under widely different conditions, in a changing environment, and for this reason they need integrators that are both unconditionally stable and that do not display unwanted algorithmic artifacts. These conditions are not always fulfilled in the existing literature, and we feel that a review of the underlying mathematical principles may be important not just for us but for other workers in the field of system biology as well.

## I. Introduction

The speed and the versatility of today's computers make the numerical modeling of complex biological systems possible, and suggest that in the future we shall be able to simulate the behavior of large cell populations *ab initio*, starting from the network of molecular reactions in single cells and climbing the ladder of complexity up to the biochemical and biophysical dynamics of whole multicellular organisms. Because of the extreme complexity of this task many existing numerical studies are limited to rather small subnets of molecular circuits within a single cell, or to the global mechanical properties of cell clusters. There are also more ambitious attempts that try to capture at least a few of the essential biochemical and biomechanical features, and [1] is a recent, comprehensive review (see also the review [2] and references therein; an incomplete list of recent references is [3-18]).

We are developing a simulation program, VBL (Virtual Biophysics Lab) that aims to include a basic description of biochemical and biomechanical features, to simulate cell clusters that should eventually be a numerical model of tumor spheroids, a useful and important *in vitro* model of solid tumors (see refs. [19-24]). In our modeling effort we proceed in a partly phenomenological way that leads to simple parameterizations; in exchange, we achieve a huge reduction in computational complexity and a considerable reduction of the space-time scale problems that affect simulations aimed at calculating the properties of macroscopic objects starting from microscopic models. We are in an advanced phase of development of the program, and we have already included cell metabolism, growth and proliferation and the extracellular environment. The 3D part of the program, i.e. geometry and biomechanical interactions, is also included, as well as a description of the intercellular microenvironment, and we can simulate large populations of dispersed cells, like those in the culture wells used for *in vitro* growth, and we have produced numerical estimates that are in good qualitative and quantitative agreement with experimental data [20, 21]. Figure 1 shows one simulated spheroid and figure 2 is an alternate display that shows the distribution of dead cells inside the same spheroid.

Although the present version of the simulation program performs well, it displays instabilities that can be traced back to the methods used to integrate the differential equations in the machinery of each individual cell and the mass-conservation equations that describe diffusion of nutrients and metabolites in the cell cluster. In this paper we discuss the generic form of the equations and ways to get rid of the instabilities: we believe that this may be a widespread problem in many systems biology projects and that this discussion may benefit other workers in the field.

## II. Differential equations in VBL

### A. Differential equations inside the cell's machinery

The simulation program includes steps that amount to the integration of a system of nonlinear differential equations [20, 21]

$$\frac{dy_k}{dt} = f_k(y_1,\ldots,y_N; x_1,\ldots,x_M; t), \qquad (1)$$

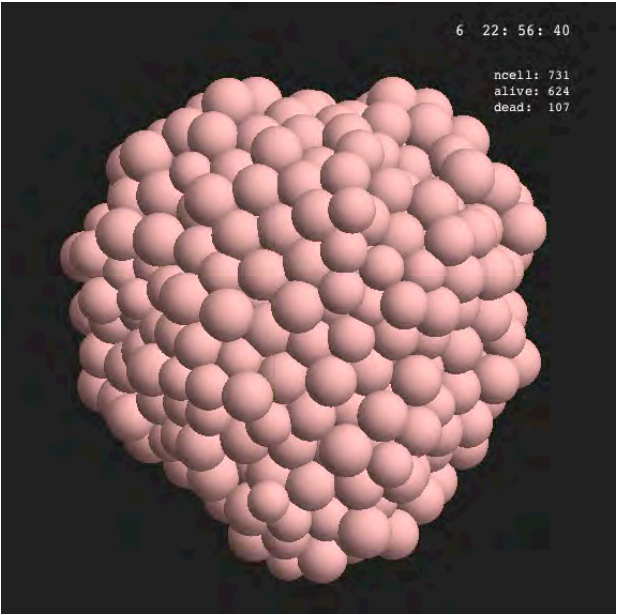

Fig. 1: Simulated tumor spheroid in VBL. The simulation started from a single cell and this cluster of cells correspond to more than 6 days of simulated time.

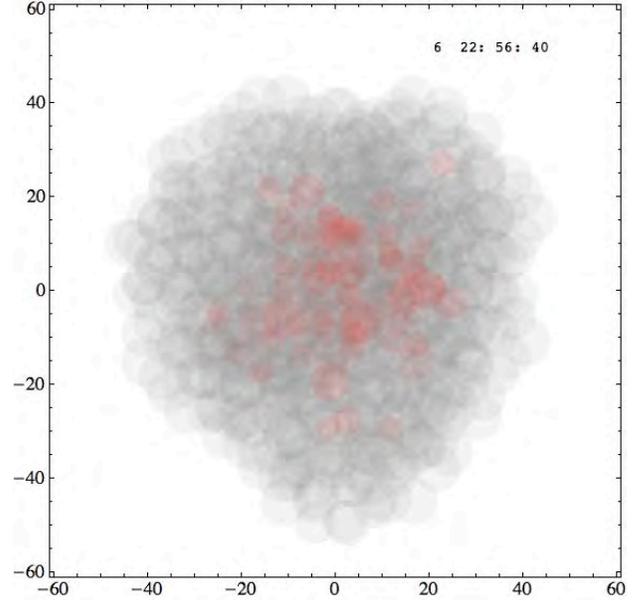

Fig. 2: Distribution of dead cells (red) inside the simulated tumor spheroid of figure 1. Here cells are represented as semitransparent balls, and dead cells in the core become visible. The numbers on the axes are µm.

where the $N$ variables $y_k$ represent the cell state, the $M$ quantities $x_l$ are additional parameters (e.g. environmental parameters), $t$ is the simulation time, and the $f_k$ functions have complex numerical dependencies. In most cases these variables represent the total mass of nutrients or metabolites in the cell. Each equation can then be recast in the time-discrete form

$$y_k^{n+1} = y_k^n + h f_k(y_1,\ldots,y_N; x_1,\ldots,x_M; t), \quad (2)$$

where $h$ is the time step and the function evaluation point is still undefined. Quite often the $f_k$ functions are evaluated at the beginning of the time interval, because in cases such as that of VBL, they are characterized by a very high complexity or they may be known only as outputs of a complicated numerical procedure, i.e.,

$$y_k^{n+1} = y_k^n + h f_k(y_1^n,\ldots,y_N^n; x_1^n,\ldots,x_M^n; nh), \quad (3)$$

so that the discretization (2) amounts to an explicit Euler integration step. It is well-known that explicit Euler integration is only conditionally stable and that it may require exceedingly small time steps; moreover even when the numerical results approach the true solution, they may display unwanted, persistent oscillations and other algorithmic artifacts.

However, in VBL the $f_k$ functions can be rewritten in the form

$$f_k = g_k(y_k) \quad (4)$$

where the $g_k$ functions depend explicitly on $y_k$ and on a set of coefficients derived from the other $y$'s and from previous history, and this means that each integration step can actually use some unconditionally stable integration method like the trapezoid method, the midpoint method, or some form of implicit Runge-Kutta. In this way the algorithmic operation of the simulated cells becomes unconditionally stable.

### B. Diffusion equations

It is not possible to review all the equations that regulate each cell in VBL in this short paper, however the diffusion equations are equally important and have similar problems, and thus in the rest of this paper we concentrate on these equations.

The diffusion equations of VBL are cast as mass conservation formulas (respectively for cells in the bulk of the spheroid, on its surface – and thus in contact with the outer environment – and for the environment)

$$V_b \frac{d\rho_b}{dt} = F(\rho_b) + D\sum_{\langle a \rangle}(\rho_a - \rho_b)g_{ab}$$

$$V_b \frac{d\rho_b}{dt} = F(\rho_b) + D\sum_{\langle a \rangle}(\rho_a - \rho_b)g_{ab} + D(\rho_A - \rho_b)g_{Ab}$$

$$V_A \frac{d\rho_A}{dt} = G(\rho_A) + D\sum_{\langle a \rangle}(\rho_a - \rho_A)g_{aA}$$

(5)

where the $\rho$'s are concentrations, $D$ is the diffusion constant, the subscripts $a$ and $b$ denote individual cells, the subscript $A$ denotes the enviroment, the $V$'s are the cells' volumes, the $g_{ab}$'s are geometric factors related to the cell volume discretization, and the notation $\langle a \rangle$ in the sum denotes the set of all cells $a$ that are adjacent to cell $b$. The $F$'s and $G$'s are slowly varying functions[1] that represent production or destruction terms related to the cells' metabolic activity, while the other terms on the rhs's represent the mass flows into and out of cells. The cells are in arbitrary positions, and not on the usual cubic lattice, so that the number of neighbors is random (in a 3D configuration it fluctuates about an average of 12).

---

[1] This means that the first time derivative of the $G$'s and $F$'s is much smaller than the largest eigenvalue of the matrix associated to the pure diffusion problem in (5).

The cell centers also move around, and cells duplicate, so that neither the positions, nor the number of equations are constant. However the mechanical motions are very slow compared to the biochemical reaction times, and cell division does not change concentrations, although it does change volumes, so that one can expect to find transients due to volume splitting in the differential system (5).

It is well known that explicit integration methods for standard diffusion equations are at best conditionally stable, but that implicit methods like Backwards Differentiation Formulas (BDF) or Crank-Nicolson (CN) are unconditionally stable (see, e.g. [25,26]). What about the unusual disordered lattice of VBL? Is it still possible to perform some stability analysis of integration methods like the standard Von Neumann stability analysis? If we neglect the mechanical rearrangement of cells, then the answer is yes. Indeed in the case of VBL, the generic equation for a cell in the bulk of the cell cluster is

$$V_b \frac{d\rho_b}{dt} = F(\rho_b) + D \sum_{\langle a \rangle} (\rho_a - \rho_b) g_{ab} \qquad (6)$$

which can be discretized to yield the difference equations

$$\rho_b^{n+1} = \rho_b^n + \frac{F(\rho_b^n) \Delta t}{V_b} + \frac{D \Delta t}{V_b} \sum_{\langle a \rangle} (\rho_a^n - \rho_b^n) g_{ab} \qquad (7)$$

(explicit integration formula, equivalent to an explicit Euler method), or

$$\rho_b^{n+1} = \rho_b^n + \frac{1}{2} \left\{ \left[ \frac{F(\rho_b^{n+1}) \Delta t}{V_b} + \frac{D \Delta t}{V_b} \sum_{\langle a \rangle} (\rho_a^{n+1} - \rho_b^{n+1}) g_{ab} \right] \right.$$
$$\left. + \left[ \frac{F(\rho_b^n) \Delta t}{V_b} + \frac{D \Delta t}{V_b} \sum_{\langle a \rangle} (\rho_a^n - \rho_b^n) g_{ab} \right] \right\} \qquad (8)$$

(Crank-Nicolson), or finally the BDF difference equations

$$\rho_b^{n+1} = \rho_b^n + \left[ \frac{F(\rho_b^{n+1}) \Delta t}{V_b} + \frac{D \Delta t}{V_b} \sum_{\langle a \rangle} (\rho_a^{n+1} - \rho_b^{n+1}) g_{ab} \right] \qquad (9)$$

Now we take test solutions $\rho_a^n \sim A^n e^{i\mathbf{k} \cdot \mathbf{r}_a}$, which are like those used in the standard Von Neumann stability analysis, but without the equal spacing assumption and neglecting the slow source terms (destruction and production), then after substitution in eq. (7) we find that the method is conditionally stable, while when we substitute in eq. (8) we find

$$A = \frac{1 - \frac{D \Delta t}{2V_b} \left[ \sum_{\langle a \rangle} g_{ab} (1 - \cos \mathbf{k} \cdot (\mathbf{r}_a - \mathbf{r}_b)) - i \sum_{\langle a \rangle} g_{ab} \sin \mathbf{k} \cdot (\mathbf{r}_a - \mathbf{r}_b) \right]}{1 + \frac{D \Delta t}{2V_b} \left[ \sum_{\langle a \rangle} g_{ab} (1 - \cos \mathbf{k} \cdot (\mathbf{r}_a - \mathbf{r}_b)) - i \sum_{\langle a \rangle} g_{ab} \sin \mathbf{k} \cdot (\mathbf{r}_a - \mathbf{r}_b) \right]} \qquad (10)$$

and likewise, after substitution in eq. (9) we find

$$A = \frac{1}{1 + \frac{D \Delta t}{V_b} \left[ \sum_{\langle a \rangle} g_{ab} (1 - \cos \mathbf{k} \cdot (\mathbf{r}_a - \mathbf{r}_b)) - i \sum_{\langle a \rangle} g_{ab} \sin \mathbf{k} \cdot (\mathbf{r}_a - \mathbf{r}_b) \right]} \qquad (11)$$

Since

$$\sum_{\langle a \rangle} g_{ab} (1 - \cos \mathbf{k} \cdot (\mathbf{r}_a - \mathbf{r}_b)) \geq 0 \qquad (12)$$

we find that in both cases (10) and (11), $|A| \leq 1$, and therefore both implicit methods are unconditionally stable on this irregular cell structure, if we neglect the slow cell movements.

III. NUMERICAL TESTS

The basic difference between the BDF and the CN methods is the order of approximation: the CN method is second order accurate in time, while the BDF method is only first order accurate [25,26]. On the other hand equation (10) means that some high frequency spatial modes have $|A| = 1$ in the CN method and this leads to spurious and sometimes long-lived oscillations [27] that are non-physical and that may mask important features of the biophysical systems. So which implicit integration method should we choose? The computational complexity is the same (in both cases we need a recursive solution of the implicit equations for the new values of the concentrations, that may be of the Jacobi or Gauss-Seidel type [26,28]), and the only actual difference may stem from the different accuracy of the two methods.

We have performed numerical tests on a simple diffusion problem described by the equation

$$\frac{\partial u}{\partial t} = D \frac{\partial^2 u}{\partial x^2} - \lambda u \qquad (13)$$

in the range $x \in [-a, a]$ with the boundary conditions $u(-a,t) = u(a,t) = u_0$ and $u(x,0) = 0$ if $|x| < a$; in this equation there is an exponential destruction term that can loosely correspond to some form of biochemical degradation. In the stationary case the equation becomes

$$D \frac{\partial^2 u}{\partial x^2} - \lambda u = 0 : \qquad (14)$$

and the stationary solution is

$$u(x, t \to \infty) = u_0 \cosh\left(\sqrt{\frac{\lambda}{D}} x\right) \Big/ \cosh\left(\sqrt{\frac{\lambda}{D}} a\right). \qquad (15)$$

For the numerical integration we have chosen parameters in a region where the oscillation problems of the CN method stand out clearly.

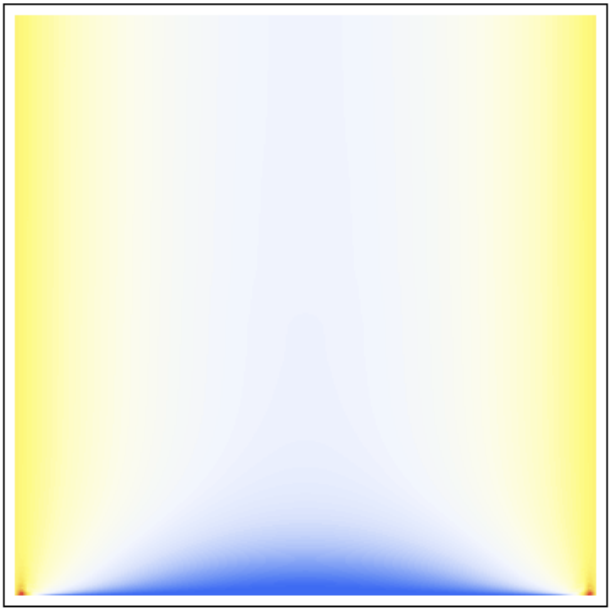

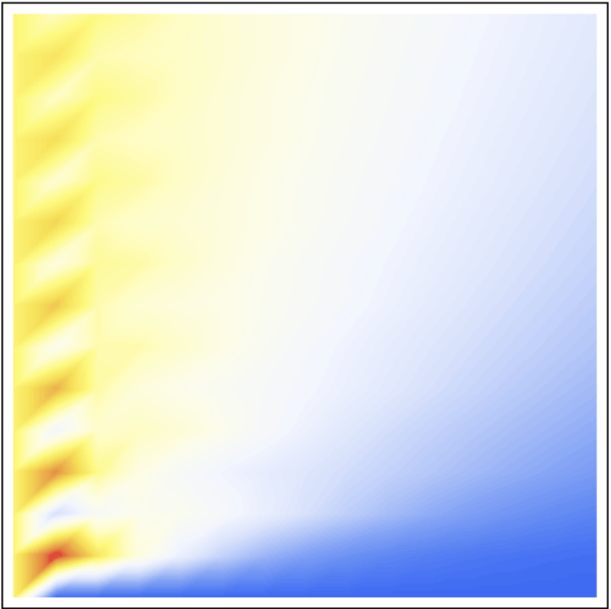

*min*                          *max*

Fig. 3: Upper part: numerical solution of equation (13) with the CN implicit method: the horizontal axis corresponds to the spatial position while the vertical axis corresponds to time, while the value of the variable *u* is mapped by color (arbitrary units). In this case the diffusion constant is rather large and there are transient oscillations that are particularly evident at the boundaries. Lower part: zoom of the bottom left corner of the upper part to show the banded structure produced by the spurious oscillations in the CN method.

Figure 3 shows the CN solution with our model parameters, while figure 4 shows the time behavior of the CN solution at a given space point, close to the boundary.

These figures have been obtained with a large diffusion constant, and in cases such as this the CN method leads to prominent oscillations near the boundaries, that in some cases last for a long time before dying out. These oscillations may show up in VBL whenever the diffusion constant is large, as is the case for oxygen. Figures 4 and 5 have been obtained with a diffusion constant 10 times larger than that in figure 3.

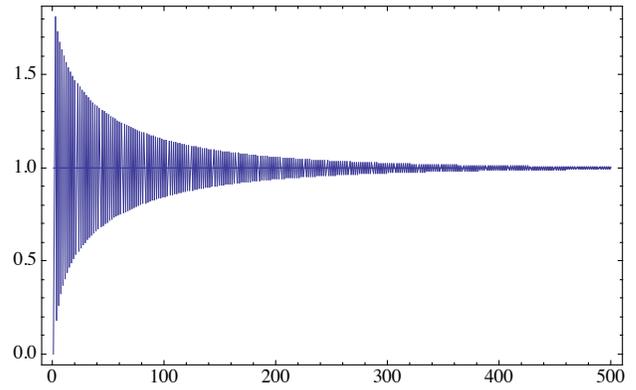

Fig. 4: Amplitude of the solution $u(x,t)$ (arbitrary units) vs. time (number of iterations) close to the boundary in the CN method. This oscillatory behavior is an artifact and it lasts for many iterations before dying out: this may adversely affect the quality of simulations.

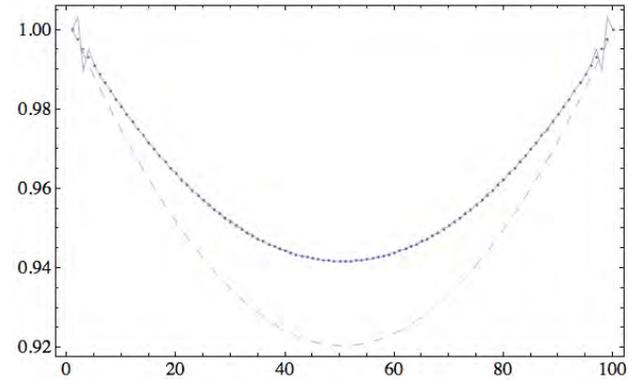

Fig. 5: Amplitude of the solution $u(x,t_{max})$ (arbitrary units) vs. space (discretized space position). The dots represent the exact solution, eq. (15), the solid line is the CN solution, and the dashed line is the BDF solution. The solutions of both methods have been obtained with the same accuracy goal. The CN solution still oscillates close to the boundary, but it is otherwise quite accurate in finding the correct solution, while the BDF solution underestimates the diffusion part of the equation.

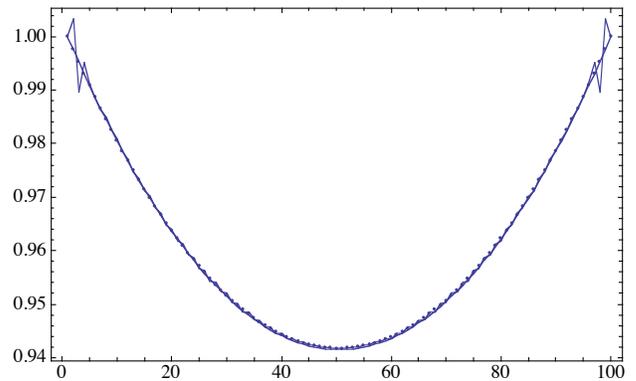

Fig. 6: Same as figure 5, but here the BDF solution has been obtained with a higher accuracy goal: both numerical solutions are very close to the exact solution, but the BDF solution does not display the annoying oscillations of the CN method. In this case computing time is approximately the same for the BDF and CN solutions even though the accuracy goals are different.

Figures 5 and 6 compare the CN and the BDF solutions at the last time step in the simulation with the exact asymptotic solution, eq. (15): while the CN solution displays unwanted oscillations near the boundaries even at this late stage, it is more accurate, while the BDF solution underestimates diffusion and leads to a reduced value of *u* at the center of the segment when the iterative solution of the implicit equation is not pushed to a high accuracy value.

The reason for the increased accuracy requirement of the BDF method is the stopping condition in the iterative solution of the implicit BDF equations (9): the iteration stops whenever two successive iterations differ less than a specified constant. Since the BDF method suppresses the high frequency spatial modes more than the CN method does (compare equations (10) and (11) ), this means that the stopping condition is reached earlier in BDF than in CN.

## IV. Conclusions

From the discussion above we can conclude that the integration steps needed inside cells are stable if we utilize implicit integration algorithms, and that the diffusion part can be handled either with the BDF method or the CN method. Although the BDF method is only first order in time, while the CN method is second order, the BDF method is certainly sufficient in contexts like that of VBL, where diffusion is always close to equilibrium.

From figures 5 and 6 we gather that the BDF method produces a high quality solution and does not display the annoying oscillations of the CM method but it may be necessary to tune its accuracy goal in the iterative solution of the implicit formula (9). Since both methods are unconditionally stable in known cases, we do not expect stability problems with either method, however, since accuracy issues may become important, and since – to the best of our knowledge – there is no comprehensive stability theory in the very general case of systems of equations such as (5) – where the parameters change, the equations change structure because of the variable number of neighboring cells, and the number of equations increases in time – it will be wise to include software switches in VBL, that allow simulations to use different implicit algorithms for the biochemical steps inside cells, and either BDF or CN for diffusion, and to compare the results obtained with different methods to exclude algorithmic artifacts.